# A NETWORK ANALYSIS APPROACH TO CONLANG RESEARCH LITERATURE

*Simon* GONZALEZ

**ABSTRACT** • The field of conlang has evidenced an important growth in the last decades. This has been the product of a wide interest in the use and study of conlangs for artistic purposes. However, one important question is what it is happening with conlang in the academic world. This paper aims to have an overall understanding of the literature on conlang research. With this we aim to give a realistic picture of the field in present days. We have implemented a computational linguistic approach, combining bibliometrics and network analysis to examine all publications available in the Scopus database. Analysing over 2300 academic publications since 1927 until 2022, we have found that Esperanto is by far the most documented conlang. Three main authors have contributed to this: Garvía R., Fiedler S., and Blanke D. The 1970s and 1980s have been the decades where the foundations of current research have been built. In terms of methodologies, language learning and experimental linguistics are the ones contributing to most to the preferred approaches of study in the field. We present the results and discuss our limitations and future work.

**1. Introduction**

The field of constructed languages (hence conlang) is a growing area of study. Early invented languages recorded have been tracked down to the 16th century (Higley, 2007), but since the origins of Esperanto in the 19[th] century, there have been great developments on different fronts, ranging from arts and philosophy (Okrand, 1985) to more logic approaches (Nicholas & Cowan, 2003). Throughout its history, conlang has been closely related to the artistic side of creation, which is observed in the widespread presence in movies and literature, for example (Campisi, 2018; Zacharias et al., 2022). On the other hand, its academic front is obtaining a more recent focus, which can be traced to approximately early 1900s with Leau & Couturat (1903) with their *Histoire de la langue universelle* presenting multiple auxiliary languages. Works like these have been crucial for the strengthening front of conlang as a field of research. With more acceptance as an academic field in the last decades, there have been great developments on linguistics that have contributed positively to our understanding of constructed languages (see Adams, 2011; Conley & Cain, 2006; Destruel, 2016; Okrent, 2010; Peterson, 2015; Rogers, 2011; Rosenfelder, 2010).

When assessing academic research on conlang, the first challenge is the terminology used. First of all, we follow the definition of conlang given by Schreyer (2021), who defines constructed languages as those consciously developed by individuals or groups, as compared to natural languages which are formed through natural processes and change over time. Schreyer also identifies four main terms used in conlang, and also specifies definitions of the terms based on the way they are created and the purpose of their creation. This offers a succinct summary of the terminology used around conlang and these terms are presented in Table 1.

| *Category* | **Term** | **Use** | **Reference** |
|---|---|---|---|
| *Full languages* | Planned languages | | Krägeloh & Neha 2014; Tonkin 2015, Gobbo 2017a |
| | Invented languages | | Lo Bianco 2004; Okrent 2009 |
| | Fictional languages | | Barnes & van Heerden 2006; Kazimierczak 2010b; Schreyer 2021b |
| | Artificial languages | | Gomez & Gerken 2000; Culbertson & Schuler 2019; Schreyer 2021a |
| *Simplified or incomplete* | Miniature artificial languages | | Kersten & Earles 2001; Carpenter 2016; Fedzechkina et al. 2016 |
| | Flavour conlangs | | Portnow 2011 |
| | Invented vocabularies | | Jackson 2011 |
| *Manner of creation* | A priori | A language made from scratch without influence from other languages | |
| | A posteriori | A language made with influences from one or more other languages | |
| *Purpose* | Auxlang | International auxiliary language | |
| | Artlang | A language used for artistic purposes, such as media or literature | |
| | Engelang | A language developed to test if something is possible in a language, often to make a "better" language than the individual's own first language | |
| | Experimental artificial languages | Languages developed by linguists and cognitive scientists to test theories of language acquisition or evolution | Christiansen 2000; Gómez & Gerken 2000; Kersten & Earles 2001; Friederici et al. 2002; Fedzechkina et al. 2016; Bartolotti & Marian 2019 |

Table 1. Adapted summary of types of conlangs as presented by Schreyer (2021)

This list shows the complexity of what conlangs are and how they are classified. This makes the search and understanding of the literature around the topic a bigger challenge.

## 2. Aim of paper

In the search of understanding the past and present of conlang research, it is vital to have a good understanding of the way the field has changed throughout the years. We aim to achieve this by carrying out a literature review through a combination of analytical techniques explained below.

Here we work under a computational/visualisation framework that allows us to understand the main body of literature that is making the most relevant contribution to the field. This, at the same time, allows us to both understand the path (past and present) of this area of study, and also analyse the seminal works that have contributed the most to our understanding and development of conlangs. These two separate tasks can be combined to create the adequate infrastructure to distinguish between studies that focus on the advancement of conlangs as a field from other studies that mention conlang keywords. By separating these, we can build stronger literature frameworks that enhance the field. This, in the end, helps us identify what are the pivotal years when the field has made the greatest progress and also the literature landmarks that have had the strongest influence.

## 3. Methodology

Identifying academic research pertinent to constructed languages is a big undertaking due to the difficulty to distinguish between works that focus on conlangs and those that refer to conlangs or treat the topic in a more tangential way. There are multiple ways to do this type of analysis, and in this paper we implement a combination of tools that can facilitate having an efficient and accurate examination of the literature available. Three tools were used based on the stage of the analysis. These are data collection by a journal API (An application programming interface), bibliometrics, and network analysis.

### 3.1. Data Collection and the Scopus API

The data was obtained through an API for the search functionality in Scopus (Elsevier) SciVal. This tool allows users to make a thorough search in academic publications using specific words within the Scopus database. An important advantage of this tool is that users can make thorough searches in all publications available regardless of the field of study and year of publication. All features available can be extracted, including abstract, keywords, references, and more. We carried out a search which aimed to capture a wide range of aspects and keywords relevant to conlang research. For the search, there was a total of 272 keywords, combining a conlang keyword (e.g., conlang, artlang, artificial), with linguistic keywords (e.g., grammar, phonology, syntax), which follows the pattern *Conlang Keyword + Linguistic Discipline*. The search identified articles and publications where the keywords appeared in any of these three sections: title, abstract and the keywords given by the authors of the publication. The Scopus database stores information on many different types of publications. For simplicity, we will refer to all observations as "publication", regardless of their type, for example, article, journal article, book, book chapter, and others.

The literature gathered through the Scopus tool has notable limitations. First, it only captures those articles and papers published in academic journals and publishing houses. It does not capture the work on conlangs outside academic publications. For example, it captures relevant literature on Esperanto, but not as much as compared to other well-known conlangs such as Elvish and Klingon. In this case, the warning is in place to point out that it does not capture all relevant works concerning conlang, but the more academic-oriented type of work. This does not mean that the nature of other non-academic type of work is not relevant, but here we focus on what has been published within the academic framework and captured in the Scopus database. Because of the nature of Scopus and the practical use of its tool to make automatic searches in its database, we believe this offers an opportunity to make searches in a wide range of academic publications and capture the majority of publications available to date.

In total, there were 2323 publications extracted from the Scopus tool, all these published from 1927 until 2022. This was the basis for the analysis of the paper. Table 2 below shows the proportion of the top ten terms searched and their results from the raw data. It shows that the term "Artificial Language" is by far the most common descriptor of conlang used in academic articles. It is important to note that the term "Constructed Language" is not as common as other terms, such as "auxiliary" "and artistic languages". Finally, other common terms are "Philosophical Language" and "Logical Language".

| *Rank* | **Search Term** | **Count** | **Proportion** |
|---|---|---|---|
| *1* | Artificial Language | 331 | 17 % |
| *2* | Artistic Language | 174 | 9.3 % |
| *3* | Philosophical Language | 172 | 9.2 % |
| *4* | Experimental Phonetics | 78 | 9.1 % |
| *5* | Logical Language | 76 | 4.1 % |
| *6* | Auxiliary Language | 70 | 4.0 % |
| *7* | Artistic Languages | 63 | 3.7 % |
| *8* | Philosophical Vocabulary | 55 | 3.3 % |
| *9* | Logical Syntax | 52 | 2.9 % |
| *10* | Constructed Language | 51 | 2.8 % |

Table 2. Rank of the top searched terms and their proportions

### 3.2. Data Processing and Natural Language Processing

The data processing and analysis was done using several scripts developed by the author in R (R Core team, 2022), using a combination of packages, mainly tidyverse (Wickham, 2019) and visNetwork (Alemende, 2021). For the data processing stage, first, we extracted the year of the publications as well as the years of the cited works in each publication. The second step was to extract the names of the authors as well as the ones that were referenced in each publication. Finally, we ran multiple Natural Language Processing (NLP) tasks to extract information from the data. NLP as a field is a branch of computer science, which trains computer algorithms to understand text and extract relevant information. It involves a combination of computational linguistics, statistical models, and machine learning. There are multiple tasks generally achieved through NLP, such as text summarization, language translation, identity extraction, among others. In this work, a first task was to extract title, abstract and keywords, from each data entry. This was a straightforward process for the keywords which were separated by punctuation marks, but it was a more difficult task for extracting and isolating relevant information from titles and abstracts. Table 3 shows the main fields of information from which relevant keywords were extracted.

| Authors | Title | Year | Keywords |
|---|---|---|---|
| *Schreyer C.* | Constructed Languages | 2021 | community of practice; constructed languages; endangered languages; Esperanto; Klingon; planned languages; speech community |
| *Martín Camacho J.C.* | On the notions of constructed language and artificial language. | 2019 | A posteriori languages; A priori languages; Artificial languages; Constructed languages; Interlinguistics |
| *Arguby Purnomo S.F.L., Nababan M., Santosa R., Kristina D.* | Ludic linguistics: A revisited taxonomy of fictional constructed language design approach for video games | 2017 | Fictional constructed language; Ludic linguistics; Taxonomy; Textonomy; Video games |
| *Sanders N.* | Constructed languages in the classroom | 2016 | Constructed languages; Creative writing; Linguistic typology; Pedagogy; Teaching linguistics |
| *Meyer A.-M.* | Slavic constructed languages in the internet age | 2016 | Constructed language; Interlinguistics; Slavic |

Table 3. Main fields analysed from the Scopus output

A first look at the languages of the original publications shows that publications were written in 34 languages. From these, English has the greatest number of publications with 1095 entries, which is a 74% of all the data collected. Russian (6%), Spanish (5%), and French (3%) make up a second group and the other languages with a smaller number of contributions, accounting for less than 2% in proportion each. In terms of the publishers, there were 357 identified ones, with the top 5 being *Cambridge University Press* with 5%, and *Routledge*, *Taylor and Francis*, *Oxford University Press*, and *Springer Netherlands* with 3% each.

### 3.3. A Timeline View of Conlang Research

Figure 1 shows a timeline of the number of publications on research about/related to conlang. It has a wide range of years, starting in 1927 until 2022. Three main stages can be observed in the figure. The first stage is from 1927 until 1975. This period is characterized by a small number of publications only for Articles/Papers and Other type of publications, which includes editorial notes and reviews, being reviews the most frequent type of document. The second observed stage is from 1975 until the beginning of the 21st century. This period is characterized by having more publications or Articles/Papers whereas the Other category remains stable. The third and final stage is from 2000 until 2022, which is a period of exponential growth, especially for the publication of articles and papers. Another important characteristic of this third stage is the appearance of books and books chapters capturing conlang research. The drop at the end of the timelines is an artifact of the data collection. Since collection was done in the early 2022, there is a decline at the end.

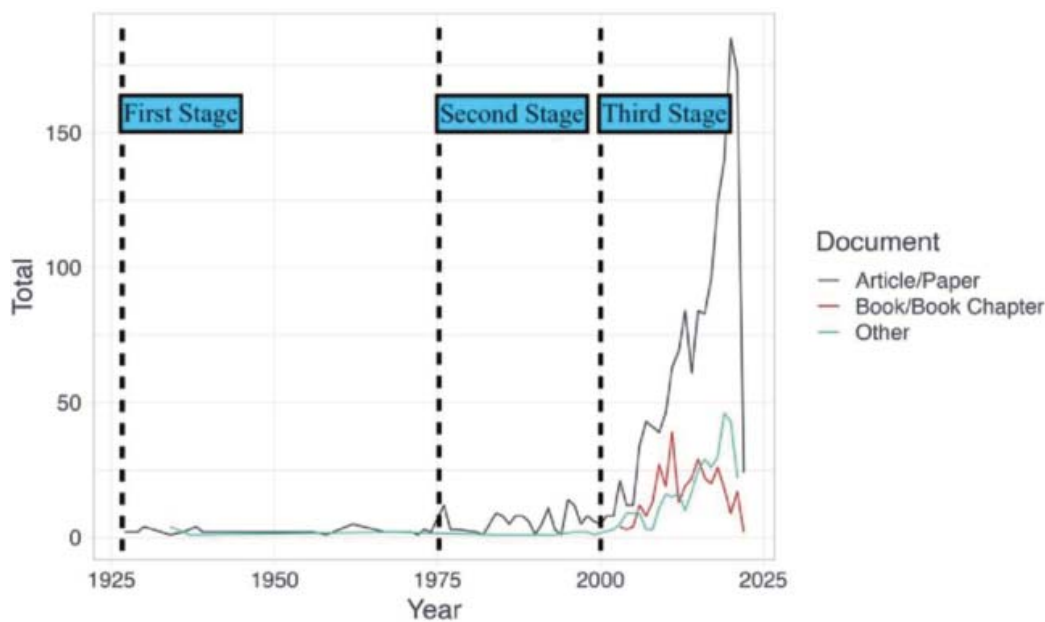

Figure 1. Timelines of all publications from the Scopus output

**4. Analysis**

One important question in this paper is how to better capture relationships between authors and pieces of work that are relevant in the topic of conlang research. This paper takes a multi-layered approach to measure these relationships. The first layer that we implement is the use of a bibliometric approach. The main purpose of this tool is to produce a quantitative evaluation of scientific published works (Shi, 2021; Szomszor et al., 2021). One of the ways to achieve it is by measuring the relationship that exists between specific publications and authors to other authors and publications (Radicchi & Castellano, 2013; Wu et al., 2022). This measurement can be done at an individual level, either on an author or a publication, or at a more general level, for example, to measure the impact that a journal has in publishing a specific author or article. The main purpose is to evaluate the impact that an article has on the topic analysed.

Since frequency (the number of times an article is cited) is generally the basis for this, we build the analysis on the number of times an article is connected to another one. Current approaches base this metric on the impact factor of a publication, for example (Dillon, 2022; Frachtenberg & McConville, 2022; Koltun & Hafner, 2021a; Koltun & Hafner, 2021b). However, we address this study from another perspective. This perspective is the second layer of the analysis and it implements a *Network Analysis* (NA) approach. This is relevant to measure relationships between authors and publications. The main advantage of NAs is that they can be used to measure the strength of relationships by observing relationships between components in a network. In this sense, NAs are commonly used to measure impact in bibliometric studies (see Cho et al., 2022; Furstenau et al., 2021; Hollebeek et al., 2022), and the focus is mainly done in measures such as h-index (M.R et al., 2021), or Citation per Paper (CPP) and Relative Citation Impact (RCI) (Sajovic & Boh Podgornik, 2022). In this paper, we measure the relationships based on the frequency of occurrence of each author and publication. Though this is not a novel approach on its own, this paper is the first one, to the best of our knowledge, that implements this approach to conlang research. Also, it is not intended to be prescriptive, but rather more descriptive. We hope that future research on this methodological approach bring more consistency to the field.

*4.1. Limitations of Current Approach*

Since the basis of the analysis is the frequency of connection between elements, we report on those connections larger than 150 connections between elements in the Years network, ten for the authors networks, and three between publication titles. This has an important impact in the results and the generalisation of the interpretation. The reason is because we are placing a hard threshold on the number of connections. This means that what the networks show is not the full picture of all the contributions because it focuses more on the ones with more frequency. This is by no means exhaustive, but it is a methodological approach that aims to be consistent in all comparisons.

Another important limitation of this methodology is on the interpretation of the network outputs. Implementing a computational approach to a growing field such as conlangs poses important challenges. One of them is that all computational outputs are interpreted based on a previous understanding of the field in which the data is found. It can be the case that two researchers looking at the same output can have different conclusions and interpret network connections differently. For this reason, all outputs are presented in full context with enumerated elements in the figures. Each output is interpreted, and the assessment is given in the text. But giving the full output of the networks also gives opportunities for readers to also assess the outputs from different perspectives.

*4.2. Understanding Network Components*

Network Analysis allows the exploration of patterns based on relationships. Figure 2 shows a representation of three basic types of networks. The first one on the left represents a centralised network. These are cases where the nodes (circles) are connected to a central node. The node sizes also represent the number of times this item appears in the data, at least in our case. The second type of network, the one in the middle, represents a dense network. In this case, the load is spread along different nodes, which shows that there is strong inter-connectivity between most or all items in the network. Finally, the figure shows a fragmented network at the right. It shows that the items group around separate data centroids. These are cases where there are clear separations between groups with individuals identifying with specific members of the same network but do not cross over. Another relevant aspect in the networks is the edges (links or lines) between nodes. Stronger links are presented by thicker lines. These show that the connections are stronger between these two nodes, compared to thinner lines which represent weaker connections. These are crucial characteristics for the assessment of the results.

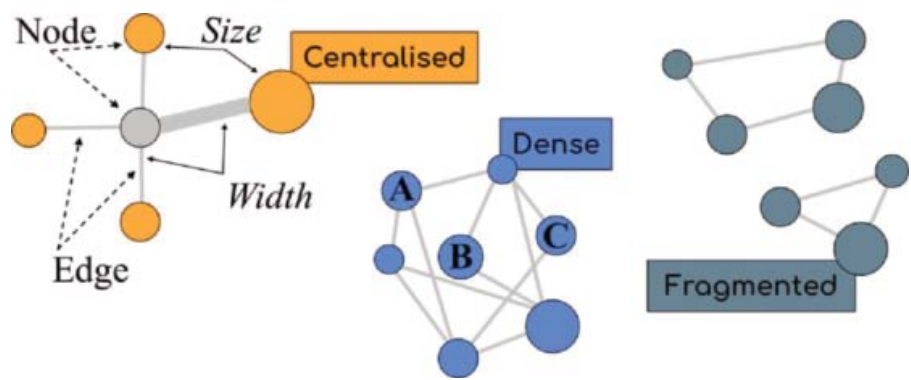

Figure 2. Network components and network types

In this paper, nodes represent either a single author, an article, or a year. The node size represents the number of times that item appears. For example, for the author's network, it represents the number of times that author appears connected to other authors in the dataset. Thus, the larger the node, the most frequent an author is in connection with other authors. The edges are the connections between nodes. Similar to size in nodes, the width of the link is also a meaningful parameter of the network. Thicker lines mean that the connections between two authors or publications is stronger (or more frequent) than thinner lines.

When assessing networks, a final important element is the proximity. Close nodes, even when they are not connected, can represent more proximity in the data. For example, in Figure 2, nodes A and B in the *Dense* network are not directly connected, but are closer to each other than nodes A and C. This is an important characteristic to evaluate, because it shows that even when two nodes are not directly connected, they are proximal to each other meaning that they can have more elements in common. In terms of clusters, networks can also show macro groups in their patterns. These can be described as an arrangement of nodes closer to each other. For example, in Figure 2, the two separate groups in the *Fragmented* network can be classified as a macro group, and this can be observed even when these are not connected through edges but are closer to each other.

## 5. Results

Results are presented in three separate networks. Each aims to examine one of the three aspects examined in this paper: authors, titles, and years.

### *5.1. Author's network*

Figure 3 shows the authors network from the data. It shows the connections between authors, which was extracted from the names of the authors in each publication, and the names of the authors in the cited works. The first assessment is that it is a strongly fragmented network. This indicates that though there are observed connections between authors, there is also strong separation between large groups. Also, there are three main macro groups identified, with the numbers for each group in blue. At the top left of the network, we see the denser macro group 1, connected by authors working on Esperanto. The macro group one shows that there are three main authors that are driving most of the traffic in relation to the works published. These authors are Garvia (A) Fiedler (B), and Blanke (C). There are direct connections between Blank (C) and Fiedler (B), and Blanke (C) and Garvia (A). There are no direct connections between Fiedler (B) and Garvia (A). However, there are works that connect Fiedler (B) and Garvia (A). Looking at individual works for these three authors (See Table 4 below), it shows that they are the ones making the stronger contributions in the field through works on Esperanto. Also, other authors are producing new network centres at the right of the macro group, but with strong connections to these three authors.

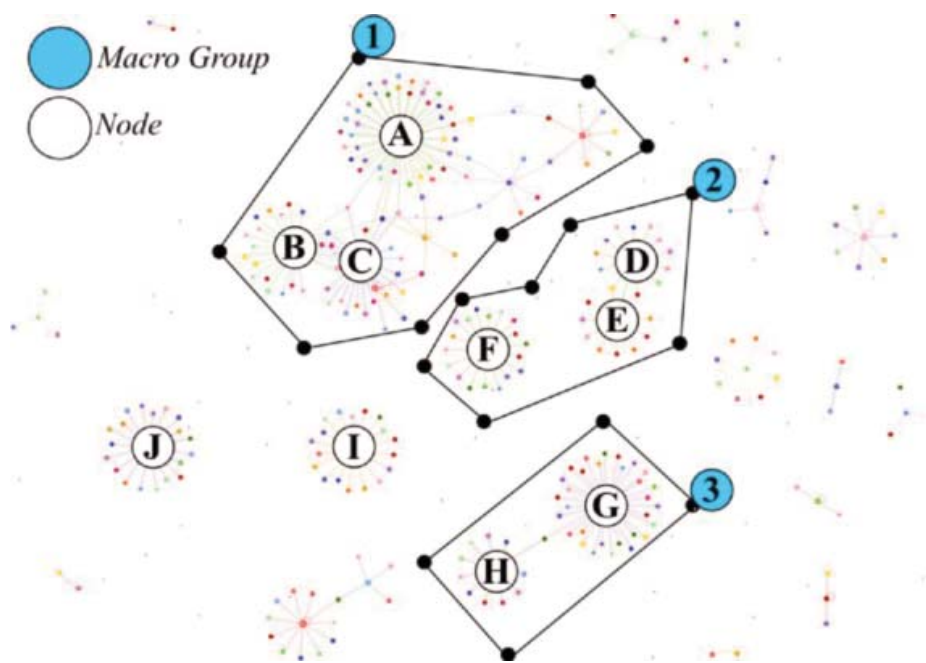

Figure 3. Author's network patterns

Looking at macro group 2, three main authors are strongly contributing to the literature. At the top right, we see Lin (D) and Finley (E) driving authors in the topic of artificial grammars and the learning component of Artificial languages. Another related topic but in a different area is driven by Muylley et al. (F), focusing more on the relationship between artificial language and secondary language learning. The final identified macro group 3 is driven by Ordin & Nespor (G) and Wiener et al. (H), who develop more work related to native languages influencing novel languages and distributional learning, focusing on the relationship between language learning and artificial languages. From this, we understand that an important part of the literature gathered focuses on the learning aspect of conlangs, which is an emerging field with strong bases for future study. This shows that language learning is positively and strongly contributing to the strengthening of conlang research. There are two other nodes at the left that are playing an important role in the network. Node I (Devitt) focuses on semantics, and node J (Brolotti & Marian) focuses on morphology-phonology. These are contributing more towards discussions on the linguistic disciplines relevant in the conlang research. The breakdown of specific authors and titles is shown in Table 4.

| *Group* | Node | Author | Selected Publications |
|---|---|---|---|
| *1* | A | Garvía | The battle of the artificial languages |
| | | | Esperanto and its rivals: The struggle for an international language |
| | | | Spelling reformers and artificial language advocates: A shifting relation |
| | | | Sotos Ochando's language movement |
| | B | Fiedler | Esperanto - a lingua franca in use: A case study on an educational NGO |
| | | | Phraseology in planned languages |
| | | | Standardization and self-regulation in an international speech community: The case of Esperanto |
| | | | The topic of planned languages (Esperanto) in the current specialist literature |
| | C | Blanke | Planned languages - a survey of some of the main problems |
| | | | Zur Rolle von Plansprachen im terminologie- wissenschaftlichen Werk von Eugen Wüster |
| | | | Interlinguistics and Esperanto studies: Paths to the scholarly literature |
| | | | Causes of the relative success of Esperanto |
| *2* | D | Lin Y.-L. | Introduction to Artificial Goals and Challenges |
| | E | Finley S. | Locality and harmony: Perspectives from artificial grammar learning |
| | | | Learning nonadjacent dependencies in phonology: Transparent vowels in vowel harmony |
| | F | Muylle M., Bernolet S., Hartsuiker R.J. | The role of L1 and L2 frequency in cross-linguistic structural priming: An artificial language learning study |
| *3* | G | Ordin M., Nespor M. | Native Language Influence in the Segmentation of a Novel Language |
| | | | Transition Probabilities and Different Levels of Prominence in Segmentation |
| | H | Wiener S., Ito K., Speer S.R. | Individual variability in the distributional learning of L2 lexical tone |
| | | | Effects of Multitalker Input and Instructional Method on the Dimension-Based Statistical Learning |
| | I | Devitt M. | Experimental Semantics |
| | J | Bartolotti J., Marian V. | Learning and processing of orthography-to-phonology mappings in a third language |
| | | | Bilinguals' Existing Languages Benefit Vocabulary Learning in a Third Language |

Table 4. Node breakdown for author and publication

### *5.2. Publication Titles Network*

The second Network shown in Figure 4 presents the connections between individual publications. Here we are looking at the interactions between these, such as references and cross citations between them. This piece of information is extracted from the references field in the Scopus output. Different from the author's network, where there was strong fragmentation, the titles network shows more connections between macro groups. The top section shows more density in the network. This indicates that there is strong cross-referencing between the ones in this section, as compared to the titles in the bottom part of the network, which displays more sparse networks. More density in networks can also be an indication that these publications above cover more mixed topics across the board. On the other hand, more sparse networks can be related to more specific publications in terms of disciplinarity, with papers having more specialised topic developments. This requires further analysis.

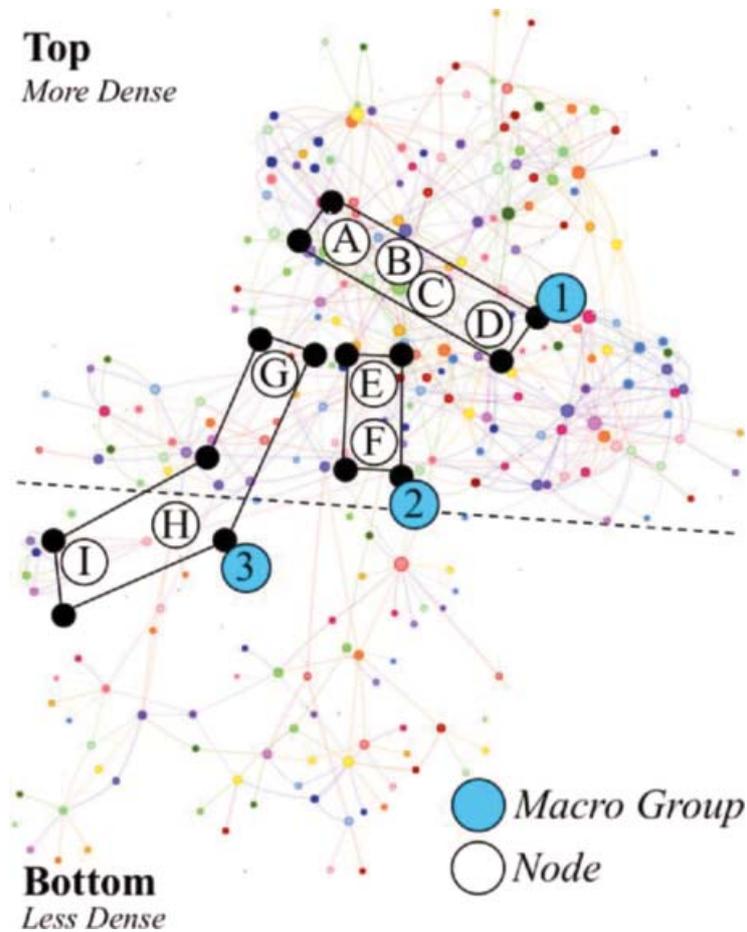

Figure 4. Title network patterns

In the dense network at the top, seven main nodes can be identified, which, in turn, can be grouped into three sections. Nodes A and B are two studies which focus on learning in children, both in statistical learning and artificial language learning. Nodes C and D are two studies which focus on the learning of artificial grammars and the artificial learning in adults and children. These nodes are grouped into macro group 1, dealing with artificial language learning in adults, children, and infants. With this type of analysis, we can examine not only the literature that is relevant, but we can also understand the areas of study that are strongly contributing to the advance of research in this field. This shows that language learning is making a positive impact on the literature.

The nodes E and F in the macro group 2, address another scientific field that is having an impact on the literature: statistics. The relevance of these two papers is that they are also showing what methodological advances have been made on this type of research, which are linear mixed effect models and mixed effects. Also, they show what are the tools being used in the method-ological approaches. The package in node E is *lme4* (Bates, 2015), which is a statistical package in R. This is an indication that this area of study is strongly influenced by social studies, looking at the social front of the field.

The third macro group touches on experiential linguistic approaches and Island Effects. The importance of island effects studies is that they bring more understanding into the syntactic phenomena in language analysis. The relevance of this in invaluable, since with this we can have a deeper understanding of how language components relate to each other, either in natural languages or artificial languages. An important observation in this network is that there are no robust connections between the statistical and methodological work nodes with the more experimental linguistic studies. This shows that there is a potential new area of study that has not been strongly explored yet. This can motivate new research on whether more work is needed in this topic. The breakdown of specific authors and titles is shown in Table 5.

| *Group* | Node | Author | Selected Publications |
|---|---|---|---|
| *1* | A | Jenny R. Saffranrichard N. Aslinand Elissa L. Newport | Statistical Learning by 8-month in infants |
| | B | Culbertson J., Schuler K. | Artificial Language Learning in Children |
| | C | Arthur S.Reber | Implicit learning of artificial grammars |
| | D | Folia V., Uddén J., De Vries M., Forkstam C., Petersson K.M. | Artificial Language Learning in Adults and Children |
| *2* | E | Douglas Bates, Martin Mächler, Ben Bolker, Steve Walker | Fitting Linear Mixed-Effects Models using Lme4 |
| | F | Dale J.Barr, Roger Levy, Christoph Scheepers, Harry J.Tily | Random Effects Structure for confirmatory hypothesis testing |
| *3* | G | Sprouse J., Hornstein N. | Experimental Syntax and Island Effects |
| | H | Devitt M. | Experimental Semantics |
| | I | Philip Hofmeister, Ivan A Sag | Cognitive constraints and Island Effects |

Table 5. Node breakdown by author and publication

### 5.3. Publication Year Network

The final network to assess is the one displaying the years of publication, and it is presented in Figure 5. Different from the two previous networks, this one displays arrow directionality. This is useful to identify in what direction the years are connected. For example, node D at the right, 1991, shows that the publications in the data from that year reference back to other papers in 1975 and 1982. Since the arrows depart from 1991 to 1975 and 1982, it shows that there are no papers cross-referencing back to 1991. As mentioned above, this does not mean that no papers have ever referenced 1991 papers, but rather, due to the threshold of 150 or more connections, any connection below the threshold was excluded. In this sense, what we have in the figure is the count of the most robust and frequent connections.

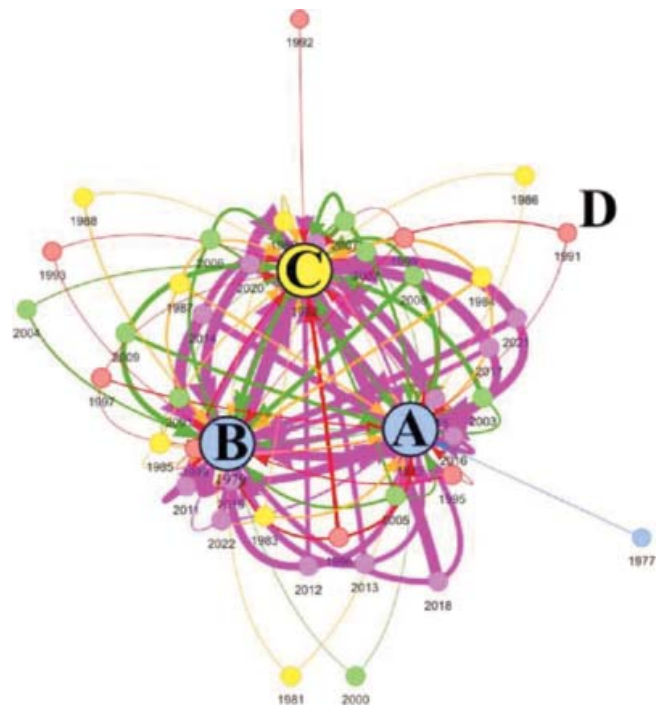

Figure 5. Year network patterns

The year was obtained from the Year column from the output. For each publication, we identified the referenced work, from which the years of publication were extracted too. In this network, each node represents a single year, and it has been colour-coded by decade. By doing this, we can observe patterns either into different decades or individual years, depending on the level of granularity desired. Similar to the top section of the *Publication Title network*, the Year network is extremely dense. Again, this shows that there is strong cross-referencing across time. This is an important finding for the field of conlangs, since it shows that as publications are produced, they are done in connections to previous research, bringing more stability and consistency in the field. We can also see the prominent role of three years driving most of the traffic across time: 1975, 1979, and 1982, or grouped into decades, the 70s and early 80s. The 90s and 2000s, though having more nodes than previous decades, are not driving traffic as compared to the three larger nodes. This indicates that there has been an increase of studies in recent years, but the majority reference back to the 70s and the 80s. This suggests that the current research foundations have been established in these two decades. The breakdown of specific authors and titles is shown in Table 6.

| *Year* | **Author** | **Publication** |
|---|---|---|
| *1975* | Ada | The Logic of Conditionals |
| | Adams & Levine | On the uncertainties transmitted from premises to conclusions in deductive inferences |
| | Agassi | Science in Flux |
| | Anderson & Belnap | Entailment: the logic of relevance and necessity |
| | Austin | How To Do Things with Words |
| | Bartlett | Beauzée's Grammaire Générale |
| | Brooks | The Well Wrought Urn: Studies in the Structure of Poetry |
| | Heinse | Ardinghello und die Glückseligen Inseln |
| *1979* | Aarts et al. | The Semantics of Adjective-noun Combinations |
| | Ackermann | Proper names, propositional attitudes and non-descriptive connotations |
| | Angenot | The Absent Paradigm: An Introduction to the Semiotics of Science Fiction |
| | Bady | Students understanding of the logic of hypothesis testing |
| | Burnley | Chaucer's Language and the Philosophers' Tradition |
| | Copeland | On when a semantics is not a semantics |
| *1982* | Popper | The Open Universe: An Argument for Indeterminism |
| | Aarsleff | From Locke to Saussure: On the order of words |
| | Anderson | Where's morphology? |
| | Ashby & Clark | Glosa newsletter |
| | Balbin | Is Esperanto an Artificial Language? |
| | Barnes | The Presocratic Philosophers Barnes: The Beliefs of a Pyrrhonist |
| | Barth & Krabbe | From Axiom to Dialogue |
| | Biagi & Luisa | Considerazioni sulla lingua di Leonardo |
| | Blanke | Plansprache und Nationalsprache |
| | Kant | Prolegomena to Any Future Metaphysics |
| | Slaughter | Universal Languages and Scientific Taxonomy in Seventeenth Century England |

Table 6. Node breakdown by author and publication

**6. Discussion**

The analysis on the available literature from the Scopus database shows that the area of conlang research is a growing field. First, there have been strong landmarks in terms of specific authors who have contributed to the field. These authors are Garvía R., Fiedler S. and Blanke D. Their stronger contribution is in the area of Esperanto. In terms of years, there have been relevant contributions since the early 1900s, but three main years are the anchors for most of the body of work. These years are 1975, 1979, and 1982. The analysis from the titles network gives information on what are the areas and subfields that are being developed in conlang as an academic field. Two major areas are language learning and experimental linguistics. Within this network we also see that among the methodologies used include random effects and mixed effects models. This is an indication that studies are looking at a multitude of factors in conlang research, which would capture the complexity of the field. Other relevant subfields include philosophical approaches and mathematical models, among other empirical methodologies. Results therefore show that this is not a homogenous field of research. Instead, it is an interdisciplinary endeavour that cuts across a wide range of disciplines, and patterns show that this is expanding.

**7. Conclusions**

Conlang research is in a decisive moment in the brink of unprecedented growth. With the implementations of emerging methodologies, the field has experienced more solidification. With a computational linguistic approach, implementing different layers of analysis, we have been able to examine the body of literature available in the Scopus database. With this, we have been able to examine to a certain depth how different authors contribute to the field and what are the main publications that have played a crucial role in the shaping of a growing field. An important point to raise is that, with the exception of Esperanto, results did not show strong research on specific conlangs, such as Klingon and Sindarin, for example. In future work, we aim to examine the research trajectory of specific conlangs. In this paper, the main purpose was to have an overall understanding of the field. To conclude, we have evidence that the field is getting more diversified in terms of methodologies, transforming thus into an interdisciplinary field with strong academic foundations.

**SIMON GONZALEZ** • is a computational linguist with a PhD on Linguistics from the University of Newcastle, Australia. Currently, he is an Honorary Lecturer at the Australian National University. He is involved in a range of projects including socio-phonetics, speech-audio forced alignment on minority languages, and computational linguistic techniques applied to different areas of interest. His recent publications include the forced alignment of an endangered language, and a machine learning implementation to the classification of socio-phonetic features in large speech corpora.